\documentclass{sig-alternate-05-2015}
\usepackage{tabulary}
\usepackage{setspace}
\usepackage{algorithm,algorithmic}
\usepackage{epsfig,graphicx,subcaption}
\usepackage[numbers,sort]{natbib}
\usepackage{rotating}
\usepackage{url}
\usepackage{float}
\usepackage{multicol,multirow}
\usepackage{booktabs}
\usepackage{amssymb,amstext,amsmath}
\usepackage[labelfont=bf]{caption}
\usepackage{color}
\usepackage{subcaption}
\usepackage{caption}
\usepackage{mathtools}
\usepackage{wrapfig}

\usepackage{subcaption} 
\usepackage{tabularx,ragged2e}

\conferenceinfo{Neu-IR '16 SIGIR Workshop on Neural Information Retrieval,}{July 21, 2016, Pisa, Italy}
\CopyrightYear{2016} 
\setcopyright{rightsretained}
\begin{document}

\title{Learning Dynamic Classes of Events using Stacked Multilayer Perceptron Networks}

\numberofauthors{3}
\author{
	\alignauthor
	Nattiya Kanhabua\\
	\affaddr{Dept. of Computer Science}\\
	\affaddr{Aalborg University, Denmark}\\
	\email{nattiya@cs.aau.dk}\\
	\alignauthor
	Huamin Ren\\
	\affaddr{Dept. of Media Technology}\\
	\affaddr{Aalborg University, Denmark}\\
	\email{hr@create.aau.dk}\\
	\alignauthor
	Thomas B. Moeslund\\
	\affaddr{Dept. of Media Technology}\\
	\affaddr{Aalborg University, Denmark}\\
	\email{tbm@create.aau.dk}\\
	}
	
\maketitle
\begin{abstract}
People often use a web search engine to find information about events of interest, for example, sport competitions, political elections, festivals and entertainment news. In this paper, we study a problem of detecting event-related queries, which is the first step before selecting a suitable time-aware retrieval model. In general, event-related information needs can be observed in query streams through various temporal patterns of user search behavior, e.g., spiky peaks for popular events, and periodicities for repetitive events. However, it is also common that users search for non-popular events, which may not exhibit temporal variations in query streams, e.g., past events recently occurred, historical events triggered by anniversaries or similar events, and future events anticipated to happen. To address the challenge of detecting dynamic classes of events, we propose a novel deep learning model to classify a given query into a predetermined set of multiple event types. Our proposed model, a Stacked Multilayer Perceptron (S-MLP) network, consists of multilayer perceptron used as a basic learning unit. We assemble stacked units to further learn complex relationships between neutrons in successive layers. To evaluate our proposed model, we conduct experiments using real-world queries and a set of manually created ground truth. Preliminary results have shown that our proposed deep learning model outperforms the state-of-the-art classification models significantly.
\end{abstract}

\section{Introduction}

Detecting event-related queries is a challenging task~\cite{Campos:2014:STI:2658850.2619088,journals/ipm/JohoJB15,INR-043}. It is the first step before applying an appropriate time-aware retrieval and ranking model, in order to improve the overall effectiveness and search experience~\cite{BerberichECIR2010,Costa:2014:LTR:2600428.2609619,DBLP:conf/ercimdl/KanhabuaN10,KanhabuaSIGIR2011c}. One way to detect events is to analyze changes in query popularity of those queries that are repeated often, rather than those in the long tail. For popular queries, search traffic typically exhibits certain temporal patterns varying over time, such as, sharp pikes, weekly- or monthly periodicity, and seasonality. Examples of spiky queries are breaking news (e.g., Syria, Zika outbreaks), celebrities (e.g., Johnny Depp, Prince), and ongoing events (e.g., tax extension, presidential candidates). Periodic or seasonal queries are, for instance, annual events (e.g., April fools' day, PGA tour) and television series (e.g., American Idol, Dancing With the Stars). However, there exist event-related queries that do not unveil temporal variations in query streams. Those queries are sporadic, which can consist of anticipated events (e.g., Samsung Note 7, Rio 2016 Olympics), and recent past or historical events (e.g., Easter, Oscars, 9/11).

\begin{figure}
	\centering
	\includegraphics[width=0.5\textwidth]{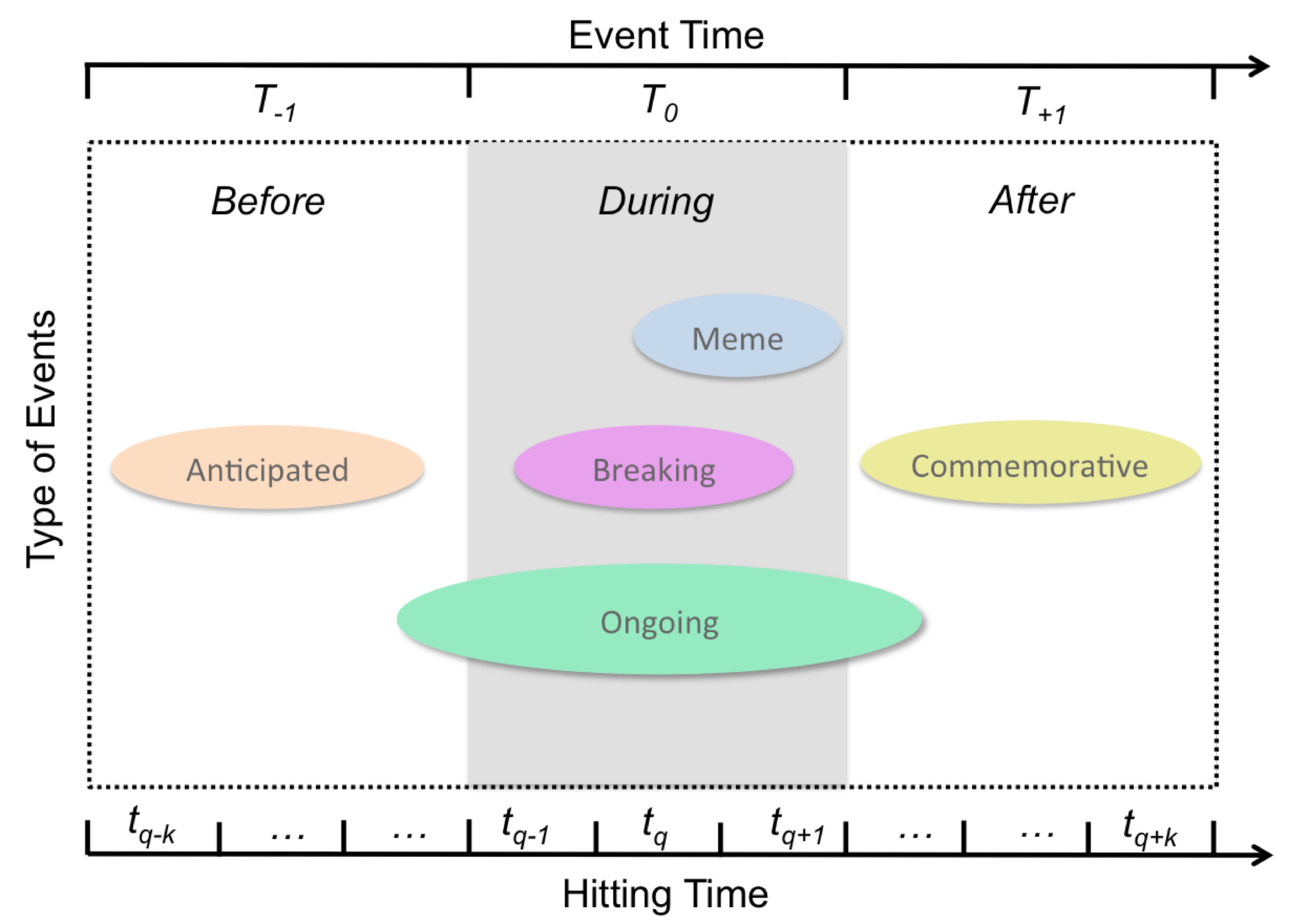}
	\caption{Dynamic Classification of Event-related Queries.}
	\label{fig:querycategory}
\end{figure}

In this paper, we address the problem of event detection in query logs that can be regarded as a classification problem. Given a set of  queries, we categorize them into the predefined dynamic classes of events. Our work differs from temporal query classification in the Temporalia Challenge~\cite{journals/ipm/JohoJB15} in at least two aspects. First, we use a novel taxonomy of more fine grained event types, which consists of six classes, i.e., \textit{anticipated}, \textit{breaking}, \textit{commemorative}, \textit{meme}, \textit{ongoing} and \textit{atemporal}. Second, we consider the time aspect of an event with respect to the time when a query was issued in order to identify its dynamic event class. In other words, the event type of a query will be dynamically determined based on the time when it was submitted (so-called \textit{hitting time}). We illustrate the dynamic event classes of queries in Figure~\ref{fig:querycategory}.

A major challenge of this task is noisy and sparse query log data. The temporal pattern of a certain type of events might not be observable because they are less popular events and often unknown in advance. Statistically, the number of queries included in each event class can be different largely; thus having impact on classification performance. A commonly used approach to handle such imbalanced data is to randomly sampling an equivalent size of training data per each event, and then to apply machine learning method on them to learn a classifier, which may lack learning ability of overall labeled data. More importantly, the learning model is not scalable to big datasets: if there will be huge amounts of queries available, the overall classification performance can be worse constrained to the event that has the minimum queries.   

We address the aforementioned challenge using a deep learning model, in which its benefits are manifold. By designing a deep framework of neuron layers, we can learn complex relationships between input data and output event classes. It is trained in an end-to-end fashion, i.e., taking instances or queries as input and providing class labels as output. Thus, deep models remove the need for manual feature engineering and have greatly reduced the need for adapting to our task. The application of deep neural networks have demonstrated their effectiveness in a wide range of fields including several challenges and benchmarks on image recognition, object detection, and many highly AI related applications, such as, image denoising, segmentation, scene understanding, speech perception and language understanding~\cite{Szegedy_2015_CVPR,Cheng_2015_CVPR,NIPS2012_4686,long_shelhamer_fcn}.  In recent years, some researches also have demonstrated the powerfulness of deep neural networks on text rankings, e.g.,~\cite{Severyn2015}.


Inspired by the fast progress of deep neural networks, and motivated by both theoretical and biological arguments, which strongly suggest that building such systems requires deep architectures and involves many layers of nonlinear processing~\cite{Bengio2009}, we propose a novel deep learning model for classifying a given query to one of these multiple events by learning a Stacked Multilayer Perceptron (S-MLP) network, which treats each multilayer perceptron as a separate learning unit and assembles either homogeneous or heterogeneous multilayer perceptron learning units to form a very deep neural network. Through this design, we learn complicated relationships between neurons in successive layers that are well expressed. Our contributions can be summarized as follows:
\begin{enumerate}
	\item We study a problem of event detection in query logs taking into account dynamic classes of events.
	\item We propose a Stacked Multilayer Perceptron (S-MLP) model, which shows the capability of processing imbalanced data from various dynamic classes of events.
	\item We conduct experiments using real-world query logs and a dataset consisting of over 10,000 instances for learning our proposed model.
\end{enumerate}

\section{Problem Statement}
Given training samples \{$\{ X_1,Y_1\},\{ X_2,Y_2\}, \dots , \{ X_m,Y_m\}$\}, each $X_i$ is a query instance, which is a vector composed of features, i.e., $X_i \in \mathbb{R}^d$, and $d$ is the dimension of feature space composed of various metrics aimed at identifying the type of events. 
We employ different features, namely, time series features from signal processing~\cite{Box:1990:TSA:574978,cleveland90, Holt2004}, along with features derived from click-through information~\cite{Dou:2007:LEA:1242572.1242651}, and standard statistical features~\cite{Jones:2007:TPQ:1247715.1247720,KleinbergKDD02}. In total, there are 28 distinct features, which are derived from two main sources (short time-span query logs and a long-term, external document collection).  The reason for leveraging these data sources is twofold. First, we want to detect events through temporal searching patterns exhibited in query logs, such as, burstiness, trend or seasonality. Second, particular event-related queries may not exhibit such temporal patterns in a query stream, but their underlying temporal information needs can be observed by analyzing the distribution of search results over time. This can compensate a lack of long time-span query logs. A list of features is summarized in Table~\ref{tab:features}. For the detailed description of each feature, please refer to the previous work~\cite{Kanhabua:2015:LDE:2740908.2741698}.

\begin{table*}[hbtp]
	\caption{List of features used in our event-related query classification task: \textit{long-span} denoting features obtained from a long time-span temporal document collection, and \textit{short-span} referring features from a query log with a short time-span.}
	\label{tab:features}
	\singlespacing
	\begin{tabulary}{1.0\textwidth}{LL|LL}
		\toprule
		\multicolumn{1}{c}{\textbf{Feature}} & \multicolumn{1}{c}{\textbf{Description}} & \multicolumn{1}{c}{\textbf{Feature}} & \multicolumn{1}{c}{\textbf{Description}}\\
		\hline
		\textit{long\_span\_acf} & long-span autocorrelation & \textit{short\_span\_acf} & short-span autocorrelation \\ 
		\textit{long\_span\_seasonal} & long-span seasonality & \textit{short\_span\_seasonal} & short-span seasonality \\
		\textit{long\_span\_kurtosis} & long-span kurtosis &  \textit{short\_span\_kurtosis} & short-span kurtosis \\
		\textit{long\_span\_KL\_PT} & long-span KL divergence & \textit{prediction\_sse} & prediction error \\
		\textit{burstLength} & longest burst duration  & \textit{t\_scope} & trending scope \\
		\textit{burstWeight} & maximum burst weight & \textit{t\_level} & trending amplitude \\
		\textit{noOfBursts} & number of bursts &  \textit{avgFreq} & average frequency\\
		\textit{isPer} & if a query contains person entities &  \textit{maxFreq} & maximum frequency\\
		\textit{isLoc} & if a query contains location entities &  \textit{CElong} & click entropy for 14 days \\
		\textit{isOrg} & if a query contains organization  entities & \textit{CEshort} & click entropy for 3 days \\
		\textit{isTempEx} & if a query contains temporal expressions& \textit{CEper} & ratio of CEshort to CElong \\
		\textit{noOfQueries} & number of queries in a query cluster (C) & \textit{sumCFreq} & sum of query frequency in C  \\
		\textit{burstDistM} & distance from the max burst & \textit{avgCFreq} & average of query frequency in C \\
		\textit{burstDistL} & distance from the longest duration burst & \textit{maxCFreq} & maximum of query frequency in C \\
		\bottomrule
	\end{tabulary}
\end{table*}

The label of each query $X_i$ is represented as $Y_i$, where 
$Y_i \in \{ 0, 1, ..., m\}$ for a multi-class classification, and $m$ is the number of classes. 
We constraint the multi-class classification problem as follows: (1)~each $Y_i$ is independent of each other, and (2)~only one event is assigned to a given query. Inspired by~\cite{DBLP:conf/icwsm/KairamMTLD13}, we propose a novel taxonomy of event-related queries consisting of \textit{anticipated} (upcoming events scheduled in a near future), \textit{breaking} (breaking news, unseen or sporadic events), \textit{meme} (viral conversation topics), \textit{commemorative} (historical events or anniversaries) and \textit{ongoing} (events being discussed or happening right now). Finally, we also consider an additional label \textit{atemporal} for cases where a given query could not be categorized as an event, or only causal behaviors from users. 

Our objective is to design a model based on deep feedforward neural networks (also called multilayer perceptrons (MLPs)), in which outputs approximate predefined target values (or ground truth) for different classes. Different from existing MLPs, we try to approximate outputs with ground truth by some approximation functions $g^*$. Each $g^*$ is a MLP unit, in another word, each $g^*$ consists of an input layer or a layer fully connected to the last layer of the previous MLP unit, one or more hidden layer(s) and an output layer of perceptrons. More specifically, given an error measurement $\delta$, the goal is to optimize: $\min \sum_{i=1}^{m}{\delta (Y_i,y_i)} $. In practice, the error between the output and the target is used to update the network parameters via the Back-propagation algorithm. We use cross-entropy as a loss function to calculate this error. Therefore, our query classification can be regarded as an optimization problem defined as: $J = \min (- \sum_{i=1}^{m}  Y_i (log (g (X_i))) )$.

\section{Stacked Multilayer Perceptron}
It has been a long held belief in the field of neural network research that the composition of several levels of nonlinearity would be key to efficiently model complex relationships between variables and to achieve better generalization performance on difficult recognition tasks~\cite{UtgoffStracuzzi02}. This viewpoint is motivated partially by knowledge of the layered architecture of regions of the human brain, such as, the visual cortex, and in part by a body of theoretical arguments in its favor. However, the problematic non-convex optimization for multilayer perceptron has for a long time prevented reaping the expected benefits~\cite{Bengio2009} of going beyond one or two hidden layers. Deep architectures~\cite{Vincent2010} have come out as a success solution at learning their parameters. Despite of its success while giving sufficient training data mainly with images, the effectiveness with limited training data is still uncertain and the design of deep architectures for IR tasks remains to be done.

\begin{figure*}
\label{fig:architecture}
\centering
\includegraphics[height=5in, width=5in,keepaspectratio]{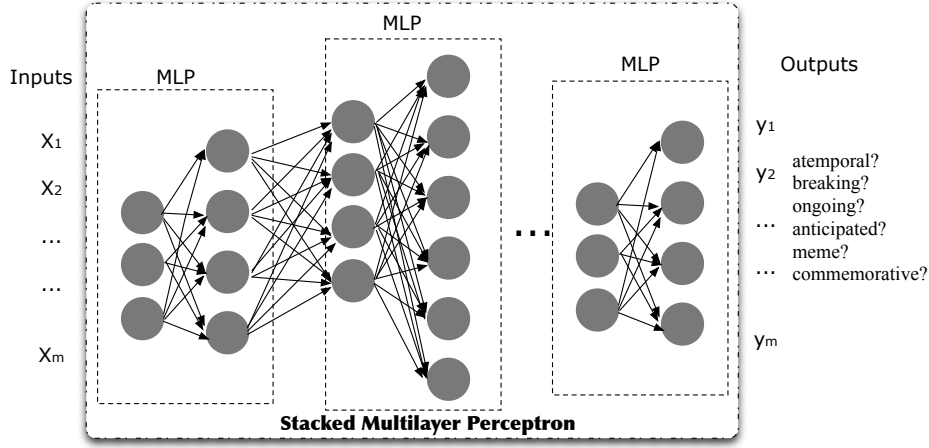}
\caption{The architecture of our proposed Stacked Multilayer Perceptron, which consists of multiple basic MLP units. Each MLP unit takes the output of the former MLP unit as input, and its output is considered as the input of successive unit. By stacking this way, complicated relationships between queries are expressive.}
\label{fig:SMLP}
\end{figure*}

Inspired by the expressiveness of multilayer perceptron, we design a novel deep neural network, namely Stacked Multilayer Perceptron (S-MLP), which uses multilayer perceptron as a basic learning unit and then assembles stacked units to learn deeper the relationships between input queries and output event. The architecture of S-MLP is illustrated in Figure~\ref{fig:SMLP}. S-MLP takes queries as input and output the event with highest probability as its label using softmax function. It is worth noting that we call each learning unit a MLP unit instead of an ordinary layer due to: 1) each MLP unit contains multiple layers and through 'stacking', thus the architecture could learn relationship between input and output even 'deeper'; and 2) each MLP unit may have a unified or non-unified structure, which means each learning block are designed with hidden layers, activation functions that can be totally the same or diverse. This design may help us understanding each MLP or the assembling of MLPs. 
%
%
%

To allow the network learn non-linear decision boundaries, we use Rectified Linear Unit (ReLU) as an activation function, i.e., $f(x)=max(0,x)$. ReLU was found to greatly accelerate (e.g., a factor of 6 in~\cite{KrizhevskySH12}) the convergence of stochastic gradient descent compared to the sigmoid/tanh functions. It is worth noting that $g^{(l)}$ has been updated $K$ iterations in the inner loop, which makes a big difference from MLP models. The consideration of updating $g^{(l)}$ is that each MLP unit can be designed either homogeneously or nonhomogeneously, i.e., the hidden layers in each MLP can either the same or diverse. Therefore, the input for each MLP unit is the output of the last layer of the preceding MLP unit. 
\section{Preliminary Results}
\label{sec:experiment}
In this section, we first explain our experimental settings including the datasets, relevance assessment, and evaluation methods. We conducted experiments on various learning methods on a MLP model, and then applied the stable learning method to our proposed S-MLP model. Learning methods play an important role while training neural networks, which affect the convergence performance, such as, speed and the capability to obtain optimization. Therefore, we varied the size of training data and compared different learning methods in their convergence rates and loss values obtained. Finally, we show comparative results with the state-of-the-art methods and discuss the experimental results of our evaluation.

\subsection{Experimental Setting}
\textbf{Query Log Datasets.} We used two real-world query log datasets publicly available. The first dataset is the AOL query logs, which consists of more than 30 million queries covering the period from March 1, to May 31, 2006.
The second dataset we used is the MSN query logs composed of about 15 million queries sampled from May 2006.
User information is anonymized and adult search queries are ignored from our study. For an external temporal collection, we used the New York Times Annotated Corpus, which contains over 1.8 million articles covering a period from January 1987 to June 2007.

There is no standard dataset for our problem of multi-class query classification. Thus, we created a set of ground-truth consisting of 837 manually identified events. We simulated hitting time, namely, every two weeks, and extracted a set of features for each event-related query. 
We asked human experts to manually classify a given query into predefined classes, while considering \textit{both} simulated hitting time \textit{and} an event time period. Particularly, an assessor gave a temporal class \textit{Label}($q$,$T_e$,$t_q$) where ($q$,$T_e$,$t_q$) is a triple of a query $q$, an event date $T_e$, and hitting time $t_q$, where each a triple was judged by at least two human assessors. 

In total, assessors evaluated 10,370 triples, which correspond to the total number of instances in our dataset available for training and testing. Finally, we obtained 988 instances of \textit{anticipated} events, 531 instances of \textit{breaking} events, 304 instances of \textit{commemorative} events, 315 instances of \textit{meme}, 2,520 instances of \textit{ongoing} events, and 5,712 instances of  \textit{atemporal} queries.	

\textbf{Methods for Comparisons.} To give a more comprehensive comparison, we divide the dataset into 70\% and 30\% as training and test; for the 70\% training, we further divide into 90\% for model learning and 10\% for validation, hence conduct cross-validation to optimize parameters on MLP model and report the effectiveness on test dataset. Note that, 70\% queries used for training are randomly selected due to the imbalance of queries per each class, training queries are also imbalanced. 

\subsection{Learning Methods of MLP}
\begin{figure}[t]
	\centering
	\includegraphics[clip, trim=1cm 1cm 1cm 1.5cm, width=0.5\textwidth]{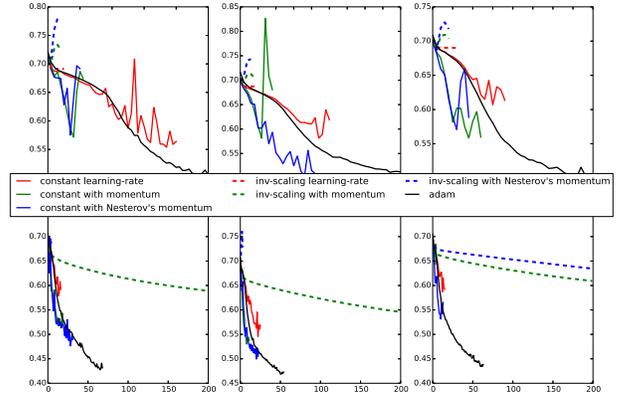}
	\caption{Comparison of various learning methods while changing the size of the training data. From top to down, left to right, the sizes of the used training data are: 20\%, 30\%, 40\%, 50\%, 60\% and 70\%. $X$ axis represents iterations, while $Y$ axis indicates loss values.}
	\label{fig:conv}
\end{figure}

Various learning methods are compared with different proportion of the training data; their convergence curves are shown in Figure~\ref{fig:conv}. As can be seen, the model tends to converge while a larger proportion of the training data is learned; while small data is used for training, any noise can lead to bias and thus there are more jitterings for most learning methods when 20\% to 40\% of training data is used. Seven types of learning methods are compared: constant learning rate, gradually decreased learning rate, gradually decreased learning rate with Nesterov's momentum, constant with momentum, gradually decreased with momentum, Adam method, and constant with Nesterov's momentum. Among them, Adam~\cite{KingmaB14} method outperforms other methods in terms of convergence and small loss function values. Thus, we adopt Adam method in the later part of experiments.


\subsection{Comparison with State-of-the-art}
\begin{figure}[t!]
\centering
\includegraphics[width=0.5\textwidth]{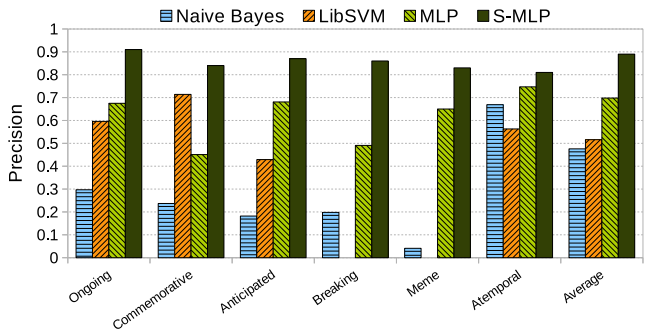}
\caption{Comparison with state-of-the-art methods on multi-class classification.}
\label{fig:stoa2}
\end{figure}

Finally, we compare our S-MLP model with state-of-the-art methods and show their comparative precision on multi-class classification in Figure~\ref{fig:stoa2}. Our S-MLP has outperformed other methods significantly, obtaining 89\% mean average precision (MAP), compared to Naive Bayes (47\%), LibSVM (51\%) and MLP (69\%). 
Remarkably, our S-MLP model outperforms other methods in classifying 'Breaking' and 'Meme' events, in which most of other methods fail. 
For a further analysis, we will perform a detailed investigation, e.g., analyzing the represented features of queries and some learning properties of our deep models. 

\section{Conclusion and Future Work}
\label{sec:conclusion}
We proposed an approach to detecting event-related queries based on a deep learning model. Deep neural networks are very expressive in representing training data, which is also verified in experiments that more MLPs are stacked in the model, the better performance is gained. Stacking multiple non-linear hidden layers makes the model very deep so that it can lean very complicated relationships between inputs and outputs, i.e., multiple classes of events. However, overfitting can be a serious problem in such networks. As pointed out in~\cite{Srivastava2014}, many of these complicated relationships will be the result of sampling noise, so they will exist in the training set but not in real test data even if it is drawn from the same distribution, which also leads to overfitting.

For our future work, over-fitting avoiding is our major consideration. One option would be to adopt units drop randomly along with their connections from the neural network during the training, which prevents unites from co-adapting too much. The neurons, which are
'dropped out' in this way, do not contribute to either the forward pass or the back propagation. So every time an input is presented, the neural network samples neurons to be dropped would lead to a different architecture. Through this, a neuron cannot rely on the presence of particular neurons, therefore, the model is forced to learn more robust features that are useful in conjunction with many different random subsets of the other neurons. To this end, it also brings us several open questions that might guide us to a better understanding of MLP unit. For example, randomly dropout is commonly used to pick neurons in one hidden layers uniformly. Can random dropout function be applied to neurons, which are located in different hidden layers yet in the same MLP unit? Will these implications function differently?



\end{document}